\documentclass[]{emulateapj}

\begin{document}
\title{
Detection of period variations in extrasolar transiting planet OGLE-TR-111b.\altaffilmark{1}
}
\author{
Rodrigo~F.~D\'{\i}az\altaffilmark{2},
Patricio~Rojo\altaffilmark{3},
Mario~Melita\altaffilmark{2},
Sergio~Hoyer\altaffilmark{3},
Dante~Minniti\altaffilmark{4,5},
Pablo~J.D.~Mauas\altaffilmark{2},
Mar\'ia~Teresa~Ru\'iz\altaffilmark{3}
}

\altaffiltext{1}{Based on observations collected with the
Very Large Telescope at Paranal Observatory (ESO Programme 278.C-5022).
}
\altaffiltext{2}{Instituto de Astronom\'{\i}a y F\'{\i}sica del
Espacio (CONICET- UBA) Buenos Aires, Argentina; rodrigo@iafe.uba.ar.}
\altaffiltext{3}{Department of Astronomy, Universidad de Chile,
Santiago, Chile.}
\altaffiltext{4}{Department of Astronomy, Pontificia Universidad
Cat\'olica, Casilla 306, Santiago 22, Chile.}
\altaffiltext{5}{Specola Vaticana, V-00120 Citta del Vaticano, Italy.}

\begin{abstract}
Two consecutive transits of planetary companion OGLE-TR-111b were
observed in the I band.  Combining these observations with data from
the literature, we find that the timing of the transits cannot be
explained by a constant period, and that the observed variations
cannot be originated by the presence of a satellite. However, a
perturbing planet with the mass of the Earth in an exterior orbit
could explain the observations if the orbit of OGLE-TR-111b is
eccentric. We also show that the eccentricity needed to explain the
observations is not ruled out by the radial velocity data found in the
literature.
\end{abstract}

\section{INTRODUCTION}
The observations of transiting extrasolar planets have produced some
of the most interesting results in the study of other planetary
systems. Their orbital configuration have permitted the first direct
measurements of radius, temperature, and composition \citep[and
references therein]{swain2008,harrington2007}, all of which are
critical to constraining the interior and evolution models of
extrasolar planets \citep[e.g.][]{fortney2008}.

 It has been further realized that the presence of variations in the
timing of transits can be attributed to otherwise undetectable planets
in the system~\citep[see, for
example,][]{miralda-escude2002,holmanmurray2005,agol2005,heylgladman2007,fordholman2007,simon2007}.
\citet{deeg2008} and \citet{ribas2008} reported indirect detections of
unseen companions by monitoring eclipse timing of the binary stellar
system \object{CM~Draconis} (1.5 M$_J$ to 0.1 M$_\odot$ candidate) and
variations in the orbital parameters of the planetary system around
\object{GJ~436} (5 $M_\oplus$ companion), respectively. However, this
last case has been recently argued against by
\citet{alonso2008}. Besides, recently-discovered transiting planets
\citep{tr182,tr211} exhibiting shifts in their radial velocities are
promising new candidates to search for variations in the timing of
their transits. On the other hand, \citet{steffenagol2005} found no
evidence of variations in the timing of transits of the
\object{TrES-1} system, after analysing data for 12 transits. Also,
after monitoring 15 transits of the star \object{HD~209458},
\citet{miller-ricci2008} were able to set tight limits to a second
planet in the system.

Here we report a significant detection of variability in the timing of
the transits of extrasolar planet \object{OGLE-TR-111b}
\citep{udalski2002,pont2004} and discuss its possible causes,
including a second unseen planet OGLE-TR-111c.

\defcitealias{tlc111}{W07}

 In a previous work \citep{minniti2007} we reported a
single transit observed in the V band which occurred around 5 minutes
before the expected time obtained using the ephemeris of
\citet[][hereafter W07]{tlc111}, but the result was inconclusive since
it had a 2.6-$\sigma$ significance.  In the present work we analyse
data of two consecutive follow-up transits of the same planet.

Section \ref{datasec} presents the new data and the reduction
procedures, in Section \ref{measusec} we describe the technique used
to measure the central times of the transits. Finally, in Section \ref{ressec}
we present our results and discuss their implications.

\section{OBSERVATIONS AND DATA REDUCTION}\label{datasec}
We observed two consecutive transits of planetary companion
OGLE-TR-111b in the I band with the FORS1 instrument at the European
Southern Observatory (ESO) Very Large Telescope (\facility VLT). The
observations were acquired during a Director's Discretionary Time run
during the nights of December 19 and December 23, 2006. Since the
orbital period of OGLE-TR-111b ($P=4.01444$ days) is almost an exact
multiple of Earth's rotational period, those were the last events
visible from the ESO facilities in Chile until May 2008.

FORS1 is a visual focal-reducer imager who had a 2048x2048 Tektronik
CCD detector and a pixel scale of 0.2 arcsec/pix. For the
observations, a nearby bright star was moved outside the field of
view, leaving OGLE-TR-111 near the center of the north-eastern
quadrant.  The chosen integration time of 6 seconds was the maximum
possible to avoid saturation of the star in case of excellent seeing.
A total of over 9 hours of observations were obtained during the
second half of both nights.  During the first night the seeing
remained stable below 0.6'', but it oscillated between 0.6'' and 1.4''
during the second night. Observations finished near local sunrise
producing a non-centered bracketing of the events and an additional
source of scatter as the sky background increased near sunrise.

We used the ISIS package \citep{alard98,alard2000} to compute precise
differential photometry with respect to a reference image in a
400$\times$400 pix sub-frame.  The reference image was obtained
combining the 10 images with best seeing, which produced an image with
$\mathrm{FWHM}\approx 0.46$ arcsec. The resulting subtracted images
were checked for abnormally large deviations or means significantly
different from zero; an image from the first night and three images
from the end of the second night were discarded in this way, leaving a
total of 488 images.

Aperture photometry was performed on the difference images using IRAF
DAOPHOT package \citep{stetson87}, which was found to give better
results than the ISIS photometry routine phot.csh \citep[see][]{hartman2004}. In
agreement with \citet{gillon2007}, we found that the scatter increased
rapidly with aperture size, although in our case the transit amplitude
remained constant (within a $0.1\%$ level). We therefore choose a
5-pixels aperture since our goal is to obtain precise measurements of
the central times of transits, and therefore the relevance of
obtaining the correct amplitude is diminished.

The uncertainty in the difference flux was estimated from the
magnitude error obtained from DAOPHOT/APPHOT, which uses Poisson
statistics, and considers the deviation in the sky background. The
flux in the reference image was measured using PSF-fitting photometry
with DAOPHOT/ALLSTARS. The systematic error introduced by this
measurement is studied further in Sect. \ref{measusec}.

\begin{figure}
\epsscale{0.95}
\centering
\plotone{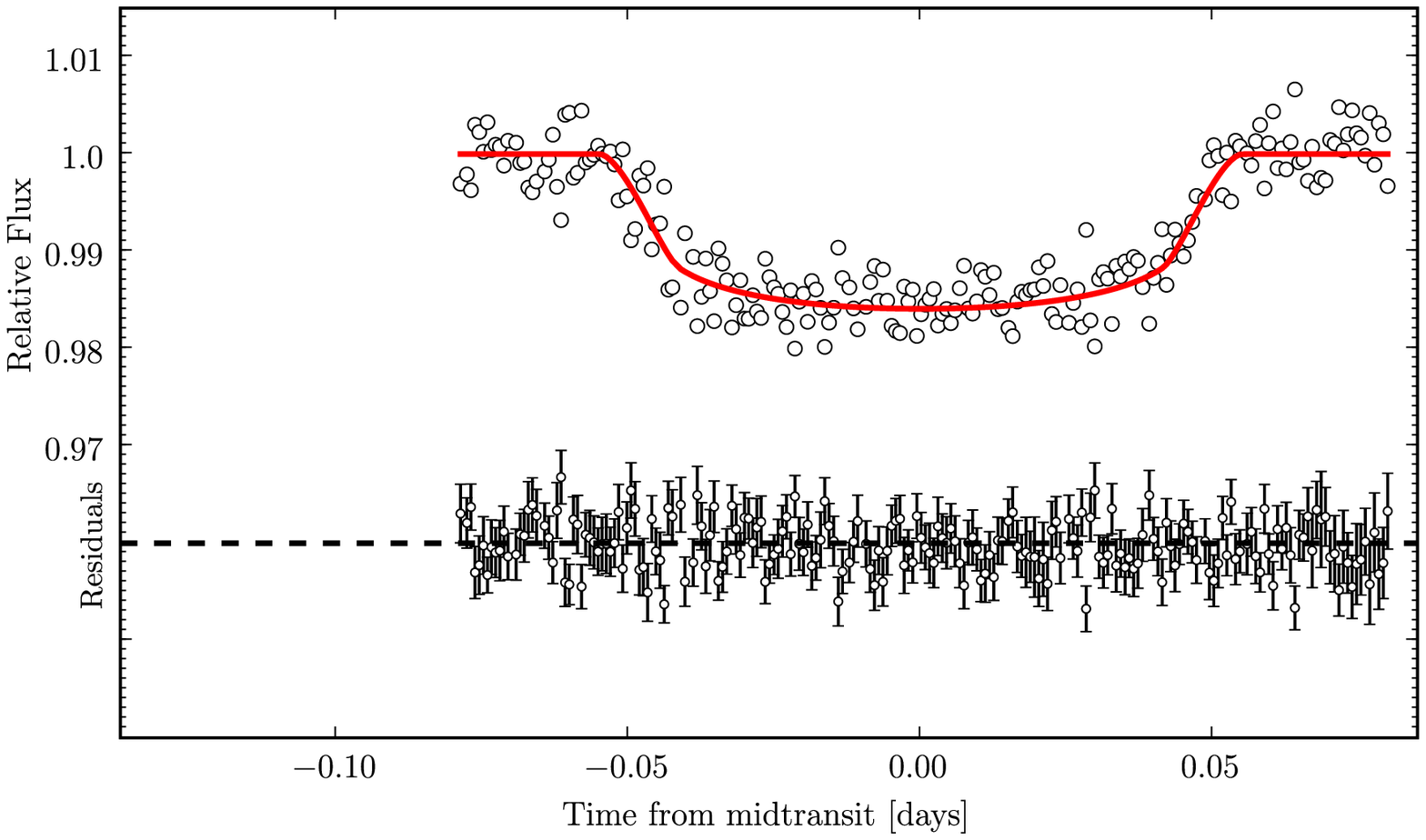}
\plotone{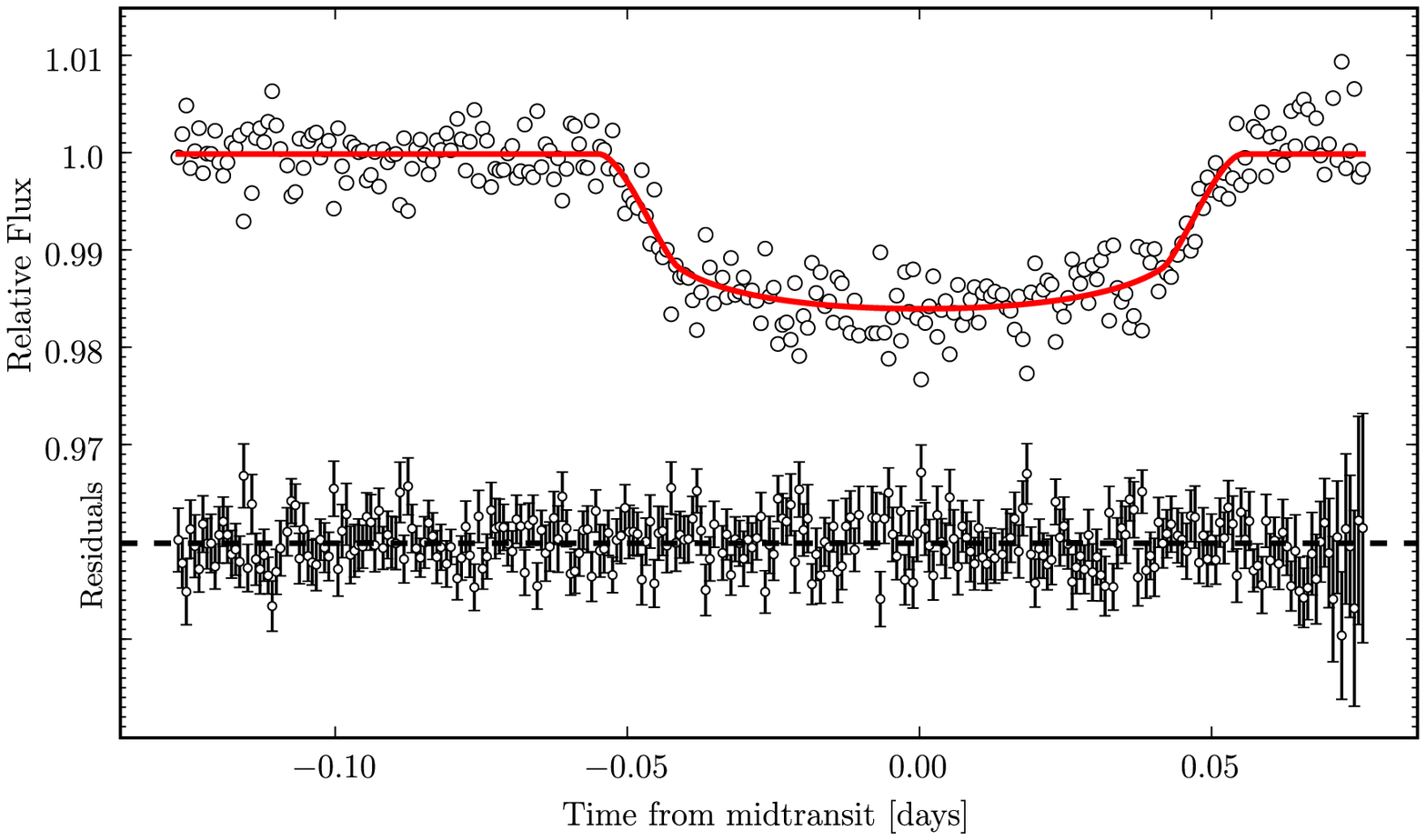}
\caption{Relative flux during two consecutive transits of planetary
  companion OGLE-TR-111b. Except for those mentioned in the text, no
  points were discarded. In the upper (lower) panel we present data
  taken on the night of December 19 (23) 2006. The residuals with the
  error bars are also shown. The dashed line
  represents the displaced zero for the residuals, and the (red) solid
  line is the best fit model. Note how the errors increase at the end of
  the second night due to the increase of the background noise caused by
  dawn.}
\label{transits}
\end{figure}

To remove possible systematics effects from the light curves we
employed the signal reconstruction method of the Trend Filtering
Algorithm \citep{tfa}. We refer readers to this paper for a
description of the method as well as for an illuminating discussion of
the possible causes of systematics effects. We chose light curves of
19 stars distributed as uniformly as possible around OGLE-TR-111 as
template light curves, and checked them for obvious variability or
uncommonly large scatter. The algorithm was iterated until the
relative difference in the curves obtained in two successive steps was
less than $10^{-5}$. The resulting science light curves for both
nights are shown in Fig.~\ref{transits}.  The standard deviation
before the transit of the second night is 2.65 mmag, almost reaching
the photon noise limit of 2.55 mmag.

\section{MEASUREMENTS}\label{measusec}
Planetary and orbital parameters, including the central times of
transits, were fitted to the OGLE-TR-111 light curve. The model used
consisted on a perfectly opaque spherical planet of radius $R_p$ and
mass $M_p$, orbiting a limb-darkened star of radius $R_s$ and mass
$M_s$ \citep{mandel2002} in a circular orbit of period $P$ and
inclination $i$. We considered a quadratic model for the
limb-darkening, with coefficients taken from \citet{claret2000} for a
star with $T_{\mathrm eff} = 5000$ K, $\log g =
4.5\;\mathrm{cm\,s^{-2}}$ and $\mathrm [Fe/H] = 0.2$ and
microturbulent velocity $\xi = 2$ km/s.  The mass of the planet and
the star were fixed to the values reported by \citet{santos2006}, $M_s
= 0.81\: M_\odot$ and $M_p = 0.52\:M_{Jup}$.  The remaining five
parameters for the model: $R_p$, $R_s$, $i$ and the central time of
each transit ($T_{c1}$ and $T_{c2}$) were adjusted using the 488 data
points of the light curve.

\begin{table}
\centering
\caption{Orbital and physical parameters for system OGLE-TR-111.}
\label{param}
\begin{tabular}{c c c}
\hline
\hline 
Parameter & Value & Confidence Limits \\
\hline
$R_s$ $[R_\odot]$& 0.811 & $_{-0.048}^{+0.041}$\\
$R_p$ $[R_{Jup}]$& 0.922 & $_{-0.067}^{+0.057}$\\
$i$ [deg] & 88.2 & $_{-0.85}^{+0.65}$\\
$t_{IV} - t_{I}$ [hr] & 2.670 & $\pm 0.014$\\
$T_{c1}$ [HJD - 2450000] & $4088.79145$ & $\pm 0.00045$ \\
$T_{c2}$ [HJD - 2450000] & $4092.80493$ & $\pm 0.00045$\\
$T_{c,VIMOS}$ [HJD - 2450000] & $3470.56389$ & $\pm 0.00055$\\
\hline
\end{tabular}
\end{table}

The parameters were obtained by minimizing the $\chi^2$ statistic
  using the downhill simplex algorithm \citep{neldermead65}
  implemented in the Scipy library\footnote{http://www.scipy.org}. We
  present the parameters in Table~\ref{param},
  and the best-fit model and the residuals in
  Fig.~\ref{transits}. Note that, except for the planetary radius and
  the time between first and last contact, the parameters reported in
  Table~\ref{param} are in agreement with previously published values
  (see Sect. \ref{ressec}).

\begin{figure}[t]
\centering
\plottwo{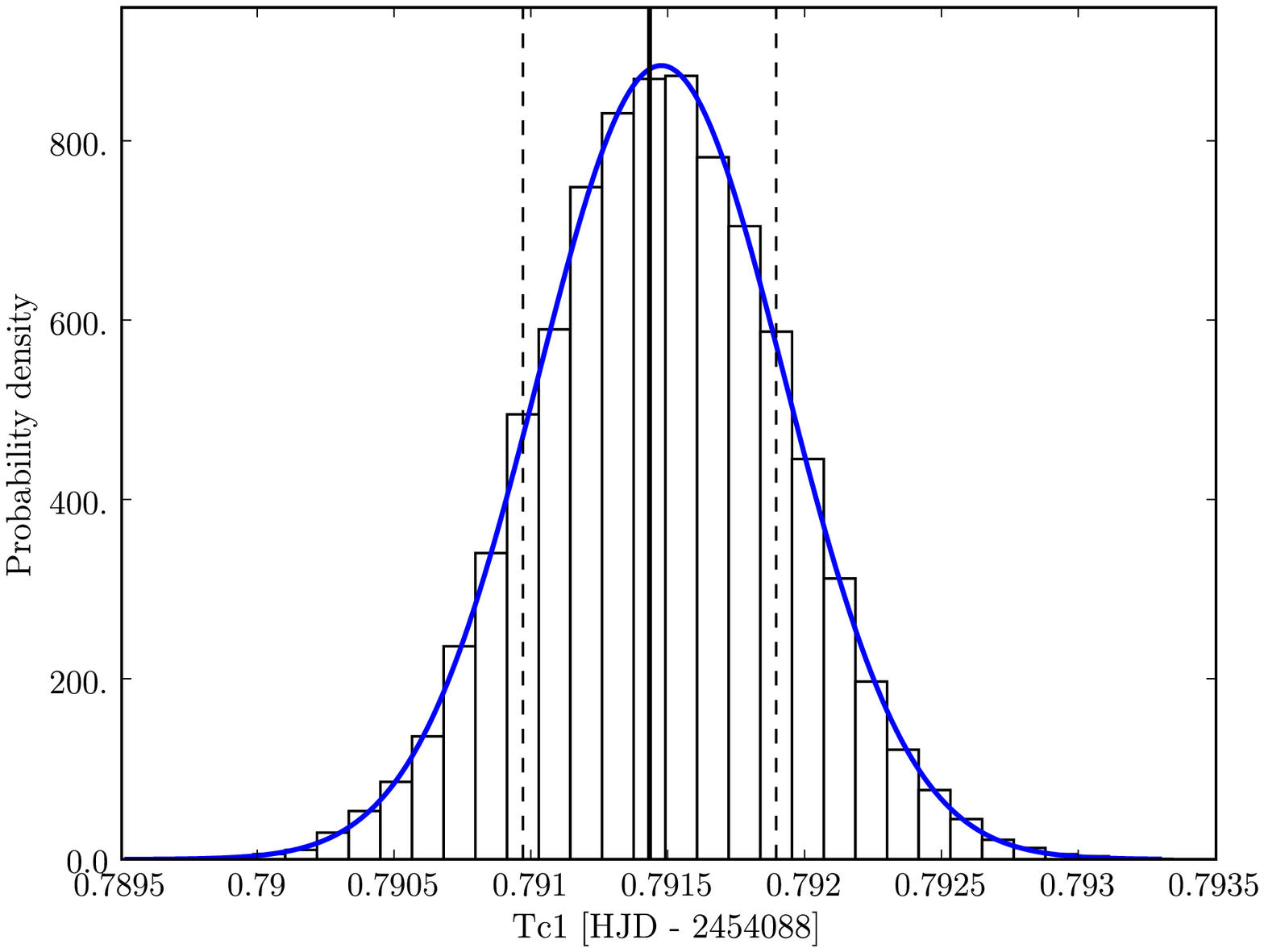}{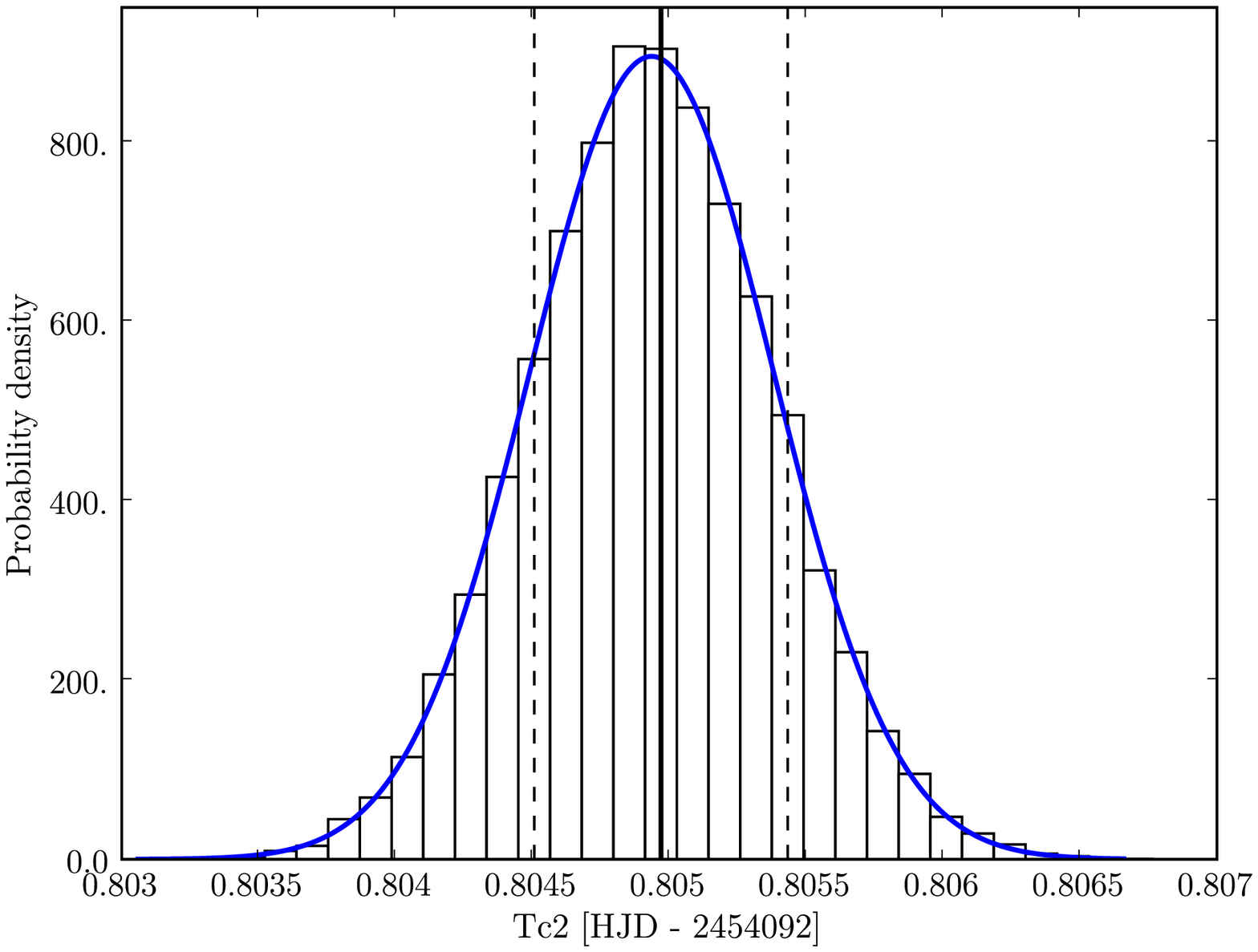}
\caption{Probability density distributions for the central times of
  the transits obtained from the MCMC simulations. The thick vertical
  solid line indicates the median of the distribution, and the dotted
  lines mark the upper and lower 68\% confidence limits. The solid
  (blue) curve is a Gaussian distribution with the same mean and
  standard deviation as the data.}
\label{hists}
\end{figure}

The uncertainties in the parameters were estimated using the Markov
Chain Monte Carlo method, which is described in detail by
\citet{tegmark2004}, \citet{ford2005} and \citet{holman2006}. We
constructed chains with 500.000 points each, and discarded the first
100.000 to guarantee convergence. The jump function employed was the
addition of a Gaussian random number to each parameter, and a global
scaling of the sigma of the random Gaussian perturbations was adjusted
after convergence was reached so that between 20\% and 30\% of the
jumps were executed.

In this manner, we built five independent chains and found that the
  mean values and the confidence intervals of the parameters (computed
  as described below) are in excellent agreement for all chains, a
  sign of good convergence.  Besides, the correlation length, defined
  as the number of steps over which the correlation function
  \citep[see][Appendix A]{tegmark2004} drops to 0.5 was about 80 for the
  central times of the transits, and around 800 for the highly
  covariant parameters $R_p$, $R_s$ and $i$, in agreement with
  \citetalias{tlc111}. This produces an effective length of about 5000
  for $T_{c1}$ and $T_{c2}$, a sign of good mixing.  

For each chain we took a random subset of 5000 values (the effective
length) of the central times and test the hypothesis that the sets
were drawn from identical populations using the Wilcoxon's rank sum
test \citep[see ][\S 14.6.9]{frodesen}. For all cases the test
statistic (which is approximately Gaussian) falls within 2.5-sigma of
the expected value, and therefore the hypothesis cannot be discarded
for significance levels below $\approx 1.2\%$.

Fig.~\ref{hists} shows two representative probability density distributions
corresponding to the two central transit times and Table~\ref{param}
reports the median and the upper and lower 68\% confidence limits,
defined in such a way that the cumulative probability below (above)
the lower (upper) confidence limit is 16\%. As a solid curve we plot
the Gaussian probability density having the same mean and standard
deviation as the data.

To test the robustness of our results, the fit was repeated fixing the
values of $R_p$, $R_s$ and $i$ to those reported by
\citetalias{tlc111} ($R_p = 1.067\:R_{Jup}$, $R_s = 0.831\:R_\odot$,
$i = 88.1$ degrees) and including the out-of-transit flux as an
adjustable parameter. The obtained times for the center of the
transits are in agreement with those reported above. The same results
are obtained if only $R_s$ is fixed to the value of
\citetalias{tlc111}.

 Additionally, to check that the systematics-removal procedure does
  not modify the shape of the light curves, we also measured the
  central times in the original curves obtained with aperture
  photometry. Again, the obtained values are in excelent agreement
  with the ones presented above, and the errors computed with MCMC are
  larger by a factor between $1.04$ and $1.99$, depending on the
  parameter, as expected.

Possible systematic errors may be introduced by the choice of the
stellar mass, the orbital period --- which affects the determination of
the orbital radius---, the model for the limb darkening, and the flux in
the reference image.  To study these effectes we obtained new fits to the data
varying the fixed parameters and the function for the limb
darkening. The stellar mass was varied by $\pm 10\%$, the photometry
in the reference image was varied by $\pm 0.1$ mag and the orbital
period by $\pm 10\,\sigma$ (see Eq.~\ref{eq:P}). The coefficients for
the quadratic limb-darkenning model were adjusted from the data
instead of fixed to the values of \citet{claret2000} and,
additionally, a linear limb darkenning model was considered, both
fixing the linear coefficient to the value computed by
\citet{claret2000} and adjusting it as part of the fit. In all cases,
the variation in the central times of transit was smaller than the
uncertainties reported in Table~\ref{param}. We therefore conclude
that the values obtained for the central transit times are robust.

\section{RESULTS AND DISCUSSION} \label{ressec}

\begin{figure}[t]
\centering
\plotone{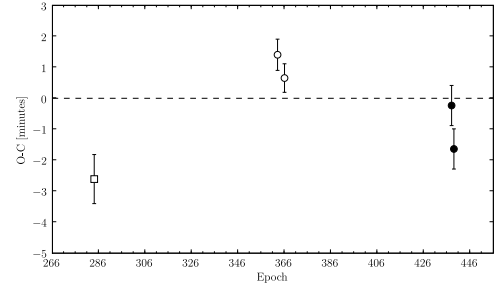}
\caption{Observed-minus-calculated times (in minutes) for the transits
of planet OGLE-TR-111b in front of its host star. The filled circles
are the new transits presented in this work, the empty circles are
from \citetalias{tlc111} and the empty square is the transit presented
by \citet{minniti2007}, which has been reprocessed for this work.}
\label{oc}
\end{figure}

We fitted a straight line to the central times of the two transits
together with those from \citetalias{tlc111} and
\citet{minniti2007}. The central time of this last transit
($T_{c,VIMOS}$) has been remeasured using the procedure described
above and the result is shown in Table~\ref{param}. In this way we
obtained a new ephemeris for the transit times:
\begin{eqnarray}
T_c &=& 2454092.80607 \pm 0.00029\; \mathrm{[HJD]}\\
P &=& 4.0144540 \pm 0.0000038\; \mathrm{days} \label{eq:P}\; \; ,
\end{eqnarray}
with correlation coefficient $\rho = 0.785$. The reduced $\chi^2$ is
9.04, indicating a poor fit. Note that the value of the period is
consistent with the value reported by \citetalias{tlc111}. The fit was
repeated including a point for the OGLE data, and
we also obtained the period from a simultaneous fit to all the
available photometry (OGLE, \citetalias{tlc111}, \citet{minniti2007}
and this work). In both cases the obtained value is in excellent
agreement with the one reported above.

In Fig.~\ref{oc} we plot the residuals of the fit. It is clear that
the observed-minus-computed (O-C) values are not consistent with a
constant period since the VIMOS transit, one of the transits from
\citetalias{tlc111} and one of the FORS transits lie -3.29-$\sigma$,
2.79-$\sigma$ and -2.52-$\sigma$ away from zero,
respectively. However, the data available to date are not enough to
determine the nature of these variations. Nevertheless, we have been
able to discard a few possibilities and study some others. We present
some preliminary results here and defer a more detailed study for a
future work.

First, the hypothesis of an exomoon seems unlikely, since the mass
needed to produce the observed O-C amplitude is at least one twenty-sixth
of the planetary mass if the moon is at a Hill radius from the
planet. However, at this distance the moon system is expected to be
unstable. For moons closer to the planet, the needed mass
increases. These are extreme values when compared with the Solar
System, where this ratio never exceeds $2.5\times 10^{-4}$
\citep{allen}.

On the other hand, several planetary system configurations reproduce
the observed trend. The equations of motion for the three-body problem
were solved with the Bulirsch-Stoer algorithm implemented in the
Mercury package \citep{chambers99} using different sets of orbital
parameters for the perturbing planet, and the results were compared
with the observations. A particularly interesting solution is that an
exterior Earth-mass planet near the 4:1 resonance produces the
observed amplitude and periodicity in the O-C times, if the orbit of
TR111b is eccentric ($e = 0.3$). On the other hand, the mass of the
perturber planet must be at least around 4 $M_{Jup}$ if the orbit of the
interior planet is nearly circular. This shows the importance of
accurately measuring the ecentricity of the interior planet through RV
data or measurements of the planet occultation
\citep[see][]{deming2007}.

\begin{figure}
\epsscale{0.95}
\centering
\plotone{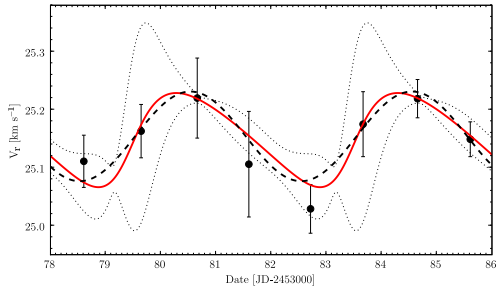}
\caption{Radial velocity measurements from \citet{pont2004} together
  with the best fit (solid line), and the corresponding $\pm 1\sigma$
  curves (dotted lines). Also shown is the fit for $e=0$ (dashed line).}
\label{rvfit}
\end{figure}

In the discovery paper by \citet{pont2004}, the orbital solution was
obtained by fixing the eccentricity of TR111b to zero. Although this
is reasonable for a single planet in a close orbit to the star, since
circularization is very effective in those conditions \citep[see, for
example,][]{zahn77}, a second planet can perturb the orbit of the
first one, increasing its eccentricity. Therefore, we reanalysed the
radial velocity (RV) data from \citet{pont2004}, in order to constrain
the possible eccentricity of the system. We found that the data are
compatible with an eccentricity of 0.3, with a reduced $\chi^2$ of
about 0.4 (for 5 degrees of freedom, see Fig.~\ref{rvfit}) compared to
the value of 0.7 for a circular orbit, as reported in the original
paper.

Additionally, note that the 1.55-$\sigma$ difference between the
  transit length presented in Table~\ref{param} and that reported by
  \citetalias{tlc111} might indicate a change in the inclination angle
  of OGLE-TR-111b \citep[see][]{ribas2008,miralda-escude2002} which
  could in principle help constrain the parameters of the perturber
  planet.

Future observations are warranted in order to pinpoint the origin of
the variation in the period of this interesting planet.
\acknowledgments
DM, PR and SH are supported by the CATA and FONDAP Center for
Astrophysics 15010003.

{\it Facilities:} \facility{VLT:Kueyen (FORS1)}


\end{document}